\documentclass[twocolumn,prb,aps,showpacs,floatfix]{revtex4}
\usepackage{graphicx}%
\usepackage{dcolumn}
\usepackage{amsmath}

\begin{document}
%\renewcommand{\thefootnote}{\fnsymbol{footnote}}
%\sloppy
%
% \newcommand{\rar}{\rightarrow}
% \newcommand{\rp}{\right)}
% \newcommand{\lp}{\left(}
% \newcommand{\Q}{{\vec q}}
\newcommand{\K}{{\vec k}}
% \newcommand{\R}{{\vec r}}
% \newcommand{\Rp}{{\vec r}^\prime}
% \newcommand{\myO}{$\ddot{\mbox{o}}$}

% spread title and abstract over two columns with \twocolumn[...text

%@twocolumnfalse\endcsname

\title{Effect of Hund's exchange on the spectral function of a
triply orbital degenerate correlated metal}

\author{P. Lombardo, A. - M. Dar\'e, R. Hayn}
\affiliation{Laboratoire Mat\'eriaux et Micro\'electronique
de Provence, CNRS UMR 6137, Universit\'e de Provence, 
B\^at. IRPHE, 49 rue Joliot-Curie, BP 146, 13384 Marseille Cedex 13 France}
\date{\today}

\begin{abstract}
We present an approach based on the dynamical mean field theory
which is able to give the excitation spectrum of a triply degenerate
Hubbard model with a Hund's exchange invariant under spin rotation. 
The lattice
problem can be mapped onto a local Anderson model containing $64$
local eigenstates. This local problem is solved by a generalized
non-crossing approximation. The influence of Hund's coupling $J$ is
examined in detail for metallic states close to the metal insulator
transition. The band-filling is shown to play a crucial role
concerning the effect of $J$ on the low energy dynamics.
\end{abstract}


\maketitle
% ] marks end of single wide column text region of
% \twocolumn[...text...]
%]

%\narrowtext

\section{Introduction}
The Mott metal insulator transition (MIT) has been the subject of many experimental
and theoretical studies over the past decades.\cite{M90,IFT98} Among correlated
materials where the Coulomb repulsion between electrons is strong enough to overcome
their kinetic energy, we find many $3d$ transition-metal oxides (see ref.~\onlinecite{M90})
and alkali-doped fullerides.\cite{G97}

The influence of orbital degeneracy on the properties of strongly
correlated electronic systems has been the subject of numerous
recent studies. It has been shown unambiguously that the orbital
degrees of freedom are of primary importance for understanding the
nature of the Mott MIT.\cite{FK97,R97,HJC98,KK96,OPB03,FG02} More
recently, the interesting and controversial possibility of an
orbital selective MIT in a multiorbital system has
been investigated.\cite{ANK02,KKR04,KKR05,L03,MGBprep05}
Hund's coupling plays an important role in these systems but
reliable dynamical quantities are not available for a spin and orbital
rotational invariant Hund's coupling. Very recently a new
approach using an extension of Wilson's numerical renormalization
group (NRG) has been proposed.\cite{PB05} The exchange part of the
Hund's coupling influences dramatically the physics
of the two-band Hubbard model and changes the nature of the Mott
transition.

In the present paper, we introduce into the triply degenerate orbital model
a non zero Hund's coupling parameter $J$ and we take into account
its full expression including the so-called spin-flip term. The only
neglected term is the orbital flip contribution or pair-hopping term between
different orbitals.\cite{DHM01}

Even the simplest lattice model for correlated electrons, the
one-band Hubbard model presents strong difficulties and
approximations are necessary to solve it. The dynamical mean field
theory (DMFT)\cite{GKK96} provides a solution in the limit of large
spatial dimension.\cite{MV89} Within the DMFT framework, a correct
description of the Mott-Hubbard metal insulator transition is
possible for the one-band Hubbard model.\cite{GKK96}

In a previous work we have proposed an approach based on the DMFT
which was able to deal with the doubly degenerate Hubbard
model\cite{LA02} with a vanishing Hund's coupling.
Metal insulator transitions for integer total
occupations per site $n=1$, $n=2$ and $n=3$ were observed, with
different values for the critical parameters on the transition
lines. Besides, the transfer of spectral weight between different
bands was interpreted in terms of microscopic dynamical processes.
Here, in order to give a more realistic description of the spectral
function of a correlated metal with several orbital degrees of
freedom close to the MIT, we propose a generalization of our previous
approach. Also, we develop a systematic description of the effects
of degeneracy, of the Hubbard interaction and the Hund's coupling 
on the triply degenerate model.

The triply degenerate model is relevant for transition-metal oxides
or other transition-metal compounds where the cubic crystal field
splits the 3$d$ level into the $e_g$ and the $t_{2g}$ manifolds
whenever it is the $t_{2g}$ manifold which is partially occupied. An
example is VO$_x$ (see ref.~\onlinecite{rata04} and
references therein). It shows a transition from metallic to
semiconducting behavior for $x$ close to 1 and the V$^{2+}$ ion
provides three electrons to the $t_{2g}$ orbitals. In that special
case, however, not only correlation but also disorder effects seem
to be very important, even for the stoichiometric samples.
Other examples are the heavy fermion compound LiV$_2$O$_4$ which remains
metallic up to very low temperature\cite{JSK99,K97} and LiTi$_2$O$_4$
which is superconducting below $T=13.7$~K.\cite{J76} 
Our approach is also relevant for MgTi$_2$O$_4$ 
(see ref.~\onlinecite{DJP05} and
references therein) which is metallic above
$T=260$~K.

In our study we investigate the $t_{2g}$-derived spectral
function. We find three different effects in the spectral function:
(i) band effects (which are usually already well described by the
local spin density approximation (LSDA)), (ii) the Hubbard
satellites (at an energy scale of $U$, already well treated by DMFT
of the one-band model), and (iii) multi-peak effects connected with
the Hund's coupling (energy scale $J$). It is the last aspect
which is new in our approach in comparison to previous studies.
Keeping in mind that these multi-peak effects should be important to
interpret realistic spectra, we analyze nevertheless a model
situation in the present work with a semi-elliptic density of states
for the uncorrelated problem.

In the following section the model Hamiltonian is presented, it
takes into account explicitly the effect of Hund's coupling.
This section also presents the application of the DMFT on this
Hamiltonian and a method of resolution based on the non-crossing
approximation (NCA). In the third section, the main results
concerning spectral properties are discussed.

\section{Hamiltonian and solving method}
The triply degenerate Hubbard Hamiltonian, including a non zero
Hund's coupling parameter $J$ can be written 
(see for instance ref.~\onlinecite{R97}) as follows~:
\begin{eqnarray}
&H&=\sum_{<i,j>,a,b,\sigma}t^{ab}_{ij}c^+_{ia\sigma} c_{jb\sigma}
+\frac{U+J}{2}\left(\sum_{i,a,\sigma} n_{ia\sigma}n_{ia-\sigma}\right) \nonumber\\
&+& \frac{U}{2}\left(\sum_{i,a\neq b,\sigma}n_{ia\sigma}n_{ib-\sigma}\right)
+ \frac{U-J}{2}\left(\sum_{i,a\neq b,\sigma}n_{ia\sigma}n_{ib\sigma}\right) \nonumber\\
&-& \frac{J}{2}\left(\sum_{i,a\neq b,\sigma}c^+_{ia\sigma}c_{ia-\sigma}
c^+_{ib-\sigma}c_{ib\sigma}\right)~. \label{hamiltonien}
\end{eqnarray}
The sum $<i,j>$ runs over
nearest neighbor sites of a Bethe lattice and $a,b=1,2,3$ are the
band indices. $c^+_{ia\sigma}$ (respectively $c_{ia\sigma}$) denotes
the creation (respectively annihilation) operator of an electron at
the lattice site $i$ with spin $\sigma$ and orbital index $a$ and
$n_{ia\sigma}$ is the occupation number operator for spin $\sigma$ and  
orbital $a$.
The hopping between
different orbitals will not be investigated here since
$t^{ab}_{ij}=-t\delta_{ab}$.
The ref.~\onlinecite{DHM01} presents a general Hamiltonian for the multi orbital
Hubbard model. The only term neglected in our approach is the pair-hopping term
between different orbitals. The prefactors in Eq.(~\ref{hamiltonien}) ensure the
Hamiltonian is invariant under any linear combination of the $t_{2g}$ orbitals.

Within the framework of the DMFT, by integrating
out all fermionic degrees of freedom
except those for a central site $i=o$, the lattice model~(\ref{hamiltonien})
can be mapped onto an effective impurity model.
The corresponding Hamiltonian $H_{\mathrm{eff}}$ contains
a local part $H_{\mathrm{loc}}$ and a part corresponding
to the coupling to the effective medium that has to be determined
self-consistently. We then write~:
\begin{equation}
   H_{\mathrm{eff}}=H_{\mathrm{loc}}+H_{\mathrm{med}}~,
   \label{model}
\end{equation}
where
\begin{eqnarray*}
 &  &H_{\mathrm{loc}}=
   \frac{U+J}{2}\left(\sum_{a,\sigma} n_{oa\sigma}n_{oa-\sigma}\right) \nonumber\\
&+& \frac{U}{2}\left(\sum_{a\neq b,\sigma}n_{oa\sigma}n_{ob-\sigma}\right)
+ \frac{U-J}{2}\left(\sum_{a\neq b,\sigma}n_{oa\sigma}n_{ob\sigma}\right) \nonumber\\
&-& \frac{J}{2}\left(\sum_{a\neq b,\sigma}c^+_{oa\sigma}c_{oa-\sigma}
c^+_{ob-\sigma}c_{ob\sigma}\right)~,
\end{eqnarray*}
and the coupling with the effective medium is
\begin{eqnarray*}
   H_{\mathrm{med}}&=&\sum_{\K a\sigma}
   \left(W_\K^a b^+_{a\K \sigma} c_{oa\sigma}+H.c.\right)\\
   &+&\sum_{\K a\sigma}\varepsilon_\K^a b^+_{a\K \sigma}b_{a\K \sigma}~.
\end{eqnarray*}
$W_\K^a$ represents the hybridization between the site $i=o$ and the
effective medium corresponding to orbital $a$. $\varepsilon_\K^a$ is
the band energy of the same effective medium. $b^+_{a\K \sigma}$
(respectively $b_{a\K \sigma}$) is the creation (respectively
annihilation) operator of an electron in the effective medium $a$.
As in the doubly degenerate situation,\cite{LA02} the effective
medium is a multi-component one that can be
characterized by the effective dynamical hybridizations~:
$$
   {\cal J}^a(\omega)=\sum_{\K}\frac{|W_\K^a|^2}{\omega+i0^+-\varepsilon_\K^a}~.
$$
By writing the equation of motion of the Hamiltonian $H_{\mathrm{eff}}$
we obtain~:
$$
   G_{a\sigma}(\omega)^{-1}=\omega-\varepsilon_o-\Sigma_{a\sigma}(\omega)
   -{\cal J}^a(\omega)~.
$$
Besides, on a Bethe lattice, the Green's function presents the
following property
$$
   G_{a\sigma}(\omega)^{-1}=\omega-\varepsilon_o-\Sigma_{a\sigma}(\omega)
   -t^2 G_{a\sigma}(\omega)~.
$$
Therefore, the self-consistent equations of the DMFT can be simply
written like
$$
{\cal J}^a(\omega)=t^2 G_{a\sigma}(\omega)~.
$$
The local problem that has to be solved self-consistently is then
formally similar to the doubly degenerate case previously investigated.
However we have to deal here with a large number $N_s$ of local impurity states $|S_i\rangle=
|\alpha_1,\alpha_2,\alpha_3\rangle$ where $\alpha_a$ represents the
electronic occupation ($0,\uparrow,\downarrow$ or $\uparrow\downarrow$) of
orbital $a$ at the impurity site $o$.
Indeed, the third orbital $a=3$ increases $N_s$ from $16$ to $64=4^3$.
Another difficulty comes from the spin-flip term~: due to this term,
the basis of local states $A=\{|S_i\rangle\}$ does no longer diagonalize $H_{\mathrm{loc}}$.

In the following, the $64$ states that diagonalize
$H_{\mathrm{loc}}$ will be noted
$|\tilde{S}_i\rangle$. They form the basis
${\cal{A}}=\{|\tilde{S}_i\rangle\}$.
Taking into account the degeneracy of local states
$|\tilde{S}_i\rangle$, it is possible to decrease the basis
dimension to $17$, the corresponding   states will be noted 
$|m\rangle$. 
Indeed the $64$ states can be classified in $13$ generic families (third column
in Table~\ref{restricted_states}).
The $k^{th}$
family contains $N_k$ local states $|\tilde{S}_i\rangle$. $g_k$ indicates
the number of different energies corresponding to this family. 
The $k^{th}$ family will be therefore represented by
$g_k$ states $|m\rangle$. We thus have $\sum_{k=1}^{13} N_k =64$ and
$\sum_{k=1}^{13} g_k =17$. (For details, see Table~\ref{restricted_states}).
%For some families, the local states $|\tilde{S}_i\rangle$ which are
%energetically degenerated are not equivalent, as we will see in the
%following by writing down the NCA equations.
%
\begin{table}[hbtp]
 \begin{center}
 \begin{tabular}{|c|c|c|c|c|c|}
 \hline \hline
 $k$ & label $i$ & $|\alpha_1,\alpha_2,\alpha_3\rangle$ & $N_k$ & $g_k$ & $|m\rangle$\\ \hline \hline
 $1$ & $1$      & $|0,0,0\rangle$           & $1$ & $1$ & $|1\rangle$ \\ \hline
 $2$ & $2$ to $7$   & $|\sigma,0,0\rangle$          & $6$ & $1$ & $|2\rangle$ \\ \hline
 $3$ & $8$ to $13$  & $|\sigma,\sigma,0\rangle$     & $6$ & $1$ & $|3\rangle$ \\ \hline
 $4$ & $14$ to $19$ & $|\sigma,-\sigma,0\rangle$        & $6$ & $2$ & $|4\rangle$,$|5\rangle$ \\ \hline
 $5$ & $20$ to $22$ & $|2,0,0\rangle$           & $3$ & $1$ & $|6\rangle$ \\ \hline
 $6$ & $23$ to $24$ & $|\sigma,\sigma,\sigma\rangle$    & $2$ & $1$ & $|7\rangle$ \\ \hline
 $7$ & $25$ to $30$ & $|\sigma,\sigma,-\sigma\rangle$   & $6$ & $3$ & $|8\rangle$,$|9\rangle$,$|10\rangle$
 \\ \hline
 $8$ & $31$ to $42$ & $|2,0,\sigma\rangle$          & $12$ & $1$    & $|11\rangle$ \\ \hline
 $9$ & $43$ to $48$ & $|\sigma,\sigma,2\rangle$     & $6$ & $1$ & $|12\rangle$ \\ \hline
 $10$ & $49$ to $54$    & $|\sigma,-\sigma,2\rangle$        & $6$ & $2$ & $|14\rangle$,$|15\rangle$
 \\  \hline
 $11$ & $55$ to $57$    & $|0,2,2\rangle$           & $3$ & $1$ & $|13\rangle$ \\ \hline
 $12$ & $58$ to $63$    & $|\sigma,2,2\rangle$          & $6$ & $1$ & $|16\rangle$ \\ \hline
 $13$ & $64$        & $|2,2,2\rangle$           & $1$ & $1$ & $|17\rangle$ \\  \hline \hline
\end{tabular}
 \caption{For each family, labels of different local states are given in 
 column $2$.
 Column $3$ displays the generic form of local state~: it stands for $N_k$
 states corresponding to $\sigma=\uparrow$ or $\downarrow$, and permutation 
 of orbitals. $g_k$ and $|m\rangle$ are defined in the text.}
 \label{restricted_states}
 \end{center}
\end{table}

Because of the spin-flip term, off-diagonal elements of
$H_{\mathrm{loc}}$ occur for local states belonging to families
$k=4, 7$ and $10$.
$k=4$ and $k=10$ families can be treated in a similar way. They can be described by
six $(2\times 2)$ blocks. As for the doubly degenerate case, singlet and triplet
states have to be considered here. For instance, in the two-dimensional subspace formed by
$|S_{14}\rangle=|\uparrow,\downarrow,0\rangle$ and $|S_{15}\rangle=|\downarrow,\uparrow,0\rangle$,
$H_{\mathrm{loc}}$ is represented by the $(2\times 2)$ non-diagonal matrix
$$
\left(
 \begin{array}{cc}
 U & -J \\
 -J & U
 \end{array}
 \right)~.
 $$
Diagonalization leads to the first eigenstate 
$|\tilde{S}_{14}\rangle=(|S_{14}\rangle+|S_{15}\rangle)/\sqrt{2}$
which corresponds to $\tilde{E}_{14}=U-J$ and
is the $S^z=0$ component of a triplet (the corresponding $S^z=\pm 1$
components are the states $|\sigma,\sigma,0\rangle$ in the $k=3$ family).
The second eigenstate is the singlet state
$|\tilde{S}_{15}\rangle=(|S_{14}\rangle-|S_{15}\rangle)/\sqrt{2}$
for which $\tilde{E}_{15}=U+J$. The notation is~: 
$|4\rangle=|\tilde{S}_{14}\rangle$
and $|5\rangle=|\tilde{S}_{15}\rangle$.

$k=7$ family is composed by two $(3\times 3)$ blocks of the following form~:
$$
\left(
 \begin{array}{ccc}
 3U-J & -J & -J\\
 -J & 3U-J & -J\\
 -J & -J   & 3U-J
 \end{array}
 \right)~.
 $$
Considering the first block where $|S_{25}\rangle=|\uparrow,\uparrow,\downarrow\rangle$,
$|S_{26}\rangle=|\uparrow,\downarrow,\uparrow\rangle$ and
$|S_{27}\rangle=|\downarrow,\uparrow,\uparrow\rangle$, we have
the eigenstates
\begin{eqnarray*}
|\tilde{S}_{25}\rangle&=&-\frac{\sqrt{6}}{3}|S_{25}\rangle+\frac{1}{\sqrt{6}}|S_{26}\rangle+\frac{1}{\sqrt{6}}
|S_{27}\rangle\\
|\tilde{S}_{26}\rangle&=&\frac{1}{\sqrt{2}}|S_{26}\rangle-\frac{1}{\sqrt{2}}|S_{27}\rangle\\
|\tilde{S}_{27}\rangle&=&\frac{1}{\sqrt{3}}|S_{25}\rangle+\frac{1}{\sqrt{3}}
|S_{26}\rangle+\frac{1}{\sqrt{3}}|S_{27}\rangle~,
\end{eqnarray*}
with the corresponding energies
\begin{eqnarray*}
\tilde{E}_{25}&=&3U\\
\tilde{E}_{26}&=&3U\\
\tilde{E}_{27}&=&3U-3J~.
\end{eqnarray*}
The state $|\tilde{S}_{27}\rangle$ plays an important role for the
dynamical properties of the triply degenerate Hubbard problem.
Indeed this low energy state is one component of the maximum
spin multiplet $S=3/2$. This fourfold multiplet is strongly stabilized
by the Hund's exchange $J$.

In table~\ref{tab_energies}, energy eigenvalues $E_m$ of $|m\rangle$ states are displayed.
\begin{table}[hbtp]
 \begin{center}
 \begin{tabular}{|c|c|c|c|}
 \hline \hline
 $k$-family & $|m\rangle$ & $n_m$  & $E_m$ \\ \hline \hline
 $1$ & $1$  & $0$   & $0$   \\  \hline
 $2$ & $2$  & $1$   & $0$   \\  \hline
 $3$ & $3$  & $2$   & $U-J$ \\  \hline
 $4$ & $4$  & $2$   & $U-J$ \\
     & $5$  & $2$   & $U+J$ \\  \hline
 $5$ & $6$  & $2$   & $U+J$ \\  \hline
 $6$ & $7$  & $3$   & $3U-3J$ \\  \hline
     & $8$  & $3$   & $3U$ \\
 $7$ & $9$  & $3$   & $3U$ \\
     & $10$ & $3$   & $3U-3J$ \\  \hline
 $8$ & $11$ & $3$   & $3U$ \\  \hline
 $9$ & $12$ & $4$   & $6U-2J$ \\  \hline
 $10$ & $14$    & $4$   & $6U-2J$ \\
      & $15$    & $4$   & $6U$ \\  \hline
 $11$ & $13$    & $4$   & $6U$ \\  \hline
 $12$ & $16$    & $5$   & $10U-2J$ \\  \hline
 $13$ & $17$    & $6$   & $15U-3J$ \\  \hline  \hline
\end{tabular}
 \caption{Energy $E_m$ of each local state $|m\rangle$. $n_m$ is the total
 on-site occupation.}
 \label{tab_energies}
 \end{center}
\end{table}

$H_{\mathrm{loc}}$ is diagonal in the basis ${\cal{A}}$, the
effective Hamiltonian  $H_{\mathrm{eff}}$ can therefore be solved by
using the extended version of the non-crossing approximation
presented in a previous work.\cite{LAS96} This approach has been
used with success for various problems.\cite{R01,RR00,SBI99,SB97}

Within NCA, propagators and self-energies for the $64$ local states
$|\tilde{S}_i\rangle$ are introduced. These propagators (and
self-energies) are identical for similar local states. For instance
$|\uparrow,\uparrow,0\rangle$, $|\uparrow,0,\uparrow\rangle$ and
$|\downarrow,\downarrow,0\rangle$ are described by the same
propagator $P_3(\omega)$ because of orbital and spin degeneracy. We
have then only $17$ different propagators $P_m(\omega)$. However,
all the $64$ local states are needed to express
the $c_{oa\sigma}$ and the $c^+_{oa\sigma}$ operators by Hubbard
operators $X_{i,j}=|\tilde{S}_i\rangle\langle\tilde{S}_j|$. More
precisely, we can write
\begin{eqnarray*}
c_{oa\sigma}&=&\sum_{i,j} D^{a,\sigma}_{i,j}|\tilde{S}_i\rangle\langle\tilde{S}_j|\\
            &=&\sum_{m,m'} {\cal D}^{a,\sigma}_{m,m'}|m\rangle\langle m'|
\end{eqnarray*}
where $D^{a,\sigma}$ is a $(64\times 64)$ matrix and ${\cal D}^{a,\sigma}$ is a
$(17\times 17)$ matrix. $D^{a,\sigma}$ is calculated by developping
the annihilation operator onto Hubbard operators and ${\cal D}^{a,\sigma}$
can be deduced by using the previously discussed degeneracy for local
propagators $P_m(\omega)$.

In the next section we present our results for the triply orbital
degenerate Hubbard model. The density of states is obtained by
solving the effective DMFT model by application of the NCA.

%%%%%%%%%%%%%%%%%%%%%%%%%%%%%%%%%%%%%%%%%%%%%%%%%%%
\section{Results and discussion}
In this section, we present the calculated densities of states for
the triply orbital degenerate Hubbard model with a Hund's exchange 
invariant under spin rotation. The DMFT results are
obtained for $t=1/\sqrt{2}$~eV, $U$ from $2$ to $6$~eV and for finite 
temperature $T = 1000$~K. We will take $J\approx U/10$ or $U/100$.

In order to understand the specific role played by Hund's coupling,
and the nature of the different bands composing the density of
states, we firstly investigate in the following sub-section the
$J=0$ situation.
The non vanishing $J$ system will be studied in section~\ref{Full_Hund_section}.

\subsection{No Hund's coupling}

\begin{figure}[hbtp]
%h=here, top, b=bottom, p=separate figure page
\begin{center}\leavevmode
\includegraphics[width=8cm]{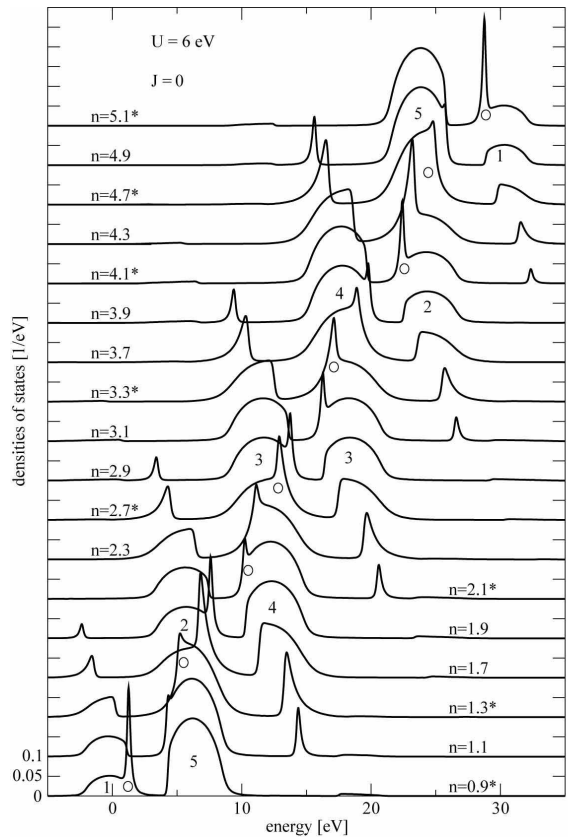}
\caption{Densities of states for various electronic occupations from
$n = 0.9$ to $n = 5.1$ for the triply degenerate Hubbard model
without Hund's coupling and for $U=6$~eV. For some fillings 
(labelled by a star), the Fermi
level is indicated by a small circle.} \label{allplots}
\end{center}
\end{figure}

Because of the triple band degeneracy, the electronic excitation
spectra present already a rich structure for $J=0$. This is shown in
figure~\ref{allplots}, where various concentrations of charge
carriers are investigated from $n=0.9$ up to $n=5.1$. In this triply
degenerate system, six electrons can in principle coexist at each
site. For some plots (labelled by a star in the figure), 
the  Fermi level position is indicated by a
small circle. These results are in agreement  with previous works on
the MIT for degenerate correlated systems. In the strongly
correlated regime, we find an insulating state for each integer
filling. For fillings close to integer values, the narrow
quasiparticle structure at the Fermi level is characteristic for a
strongly correlated metal.
With increasing chemical potential, an important transfer of
spectral weight is observed. The relative band spectral weights can
be interpreted in terms of probabilities. For instance, for $n=0.9$
the excitation spectrum is mainly composed of two bands
corresponding to the lower and the upper Hubbard bands (LHB and
UHB). The number of occupied electronic states of the LHB gives the
probability of removing an electron from the system, like in a
photoemission experiment. The spectral weight of the UHB gives the
probability of adding an electron in an empty state 
(inverse photoemission). For $n$ close
to $1$ there is only one possibility for removing an electron, but
there are five possibilities of adding an electron onto the same
site. This is indicated in figure~\ref{allplots} for various densities~:
$(1,5)$, $(2,4)$\ldots  For $J=0$, the additional energy involved in this process is
$U$, whatever the spin of the electron and the occupied orbital are.
%%%%%%%%%%%%%%%%%%%%%%%%%%%%%%%%%%%%%%%%%%
%%%%%%%%%%%%%%%%%%%%%%%%%%%%%%%%%%%%%%%%%%%%%%%

\begin{figure}[hbtp]
%h=here, t=top, b=bottom, p=separate figure page
\begin{center}\leavevmode
\includegraphics[width=8cm]{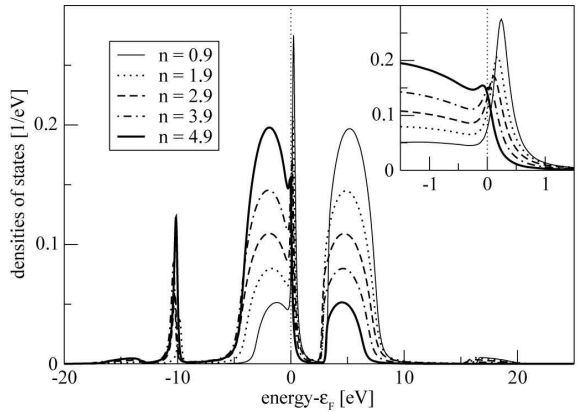}
\caption{Densities of states for various electronic occupations
without Hund's coupling. For each plot $U=6$~eV and the Fermi level
corresponds to the origin of the graph. The inset shows the low
energy behavior.} \label{allplots_ef}
\end{center}
\end{figure}

Densities of states for $n=0.9, 1.9, 2.9, 3.9$ and $4.9$ are
displayed in figure~\ref{allplots_ef} with an energy shift ensuring
that all Fermi energies are at the same point. Here, the transfer of
spectral weight from the LHB to the UHB is obvious. The inset shows
the low energy behavior. We do not consider exactly integer values for $n$
to prevent numerical problems when the density 
of states at the Fermi level is very close to zero. Nevertheless, our approach is 
numerically well defined for filling arbitrary close to integer 
values, with increasing numerical efforts. For all fillings, 
a quasiparticle peak is
clearly present, but the effective mass of these quasiparticles is
strongly dependent on the number of charge carriers. The correlated
system with $n = 0.9$ is much closer to the insulating state than
the $n = 4.9$ situation. The $n=0.9$ effective mass can be
obtained from the self-energy via $m^*/m=\left. (1-\partial {\rm Re}\Sigma(\omega)/\partial \omega) \right|_{\omega=0}$ and is approximately
seven times the $n=4.9$ effective mass. Actually, because of
electron-hole symmetry, $n = 1.1$ and $n = 4.9$ give symmetric
densities of states and therefore the same effective masses for
quasiparticles. Consequently, starting from $n = 1$, a $10\%$
electron or hole doping leads to very different systems contrasting
with the one band Hubbard model.
%%%%%%%%%%%%%%%%%%%%%%%%%%%%%%%%%%%%%%%%%%
%%%%%%%%%%%%%%%%%%%%%%%%%%%%%%%%%%%%%%%%%%%%%%%
For $J=0$, we obtained a Mott metal
insulator transition for each integer filling $n=1,2,3,4,5$.
This is consistent with previous theoretical works using slave-boson
formalism,\cite{L94,H97,FK97,KS98}
variational method\cite{BW97} and quantum Monte-Carlo dynamical mean field theory.\cite{R97,HJC98}
Nevertheless, the interesting question of the order of the transition
cannot be investigated by the present approach because of the  
NCA pathology
at low temperature.\cite{PCJ93,M84} Indeed, the coexistence 
region extending to finite
doping off half-filing observed for the doubly degenerate Hubbard
model has been observed only at low temperature.\cite{FL02,BWG98}

\subsection{Non vanishing Hund's  coupling}
\label{Full_Hund_section}

Keeping a spin-rotational invariant Hund's exchange means that we have
to handle explicitly the last term in expression~(\ref{hamiltonien}).
Recently, a very accurate study showing the importance of this term has
been presented for the two-band Hubbard model.\cite{PB05} Indeed, a
completely different behavior occurs if one replaces the rotationally
invariant Hund exchange by an Ising-like one.

Here we will focus on the specific effects of the Hund's
exchange coupling on the electron dynamics for the strongly
correlated metal and for the integer filling system which is
insulating for large $U$. We will see
that $n$ plays a crucial role.

%%%%%%%%%%%%%%%%%%%%%%%%%%%%%%%%%%%%%%%%%%
\begin{figure}[hbtp]
%h=here, t=top, b=bottom, p=separate figure page
\begin{center}\leavevmode
\includegraphics[width=8cm]{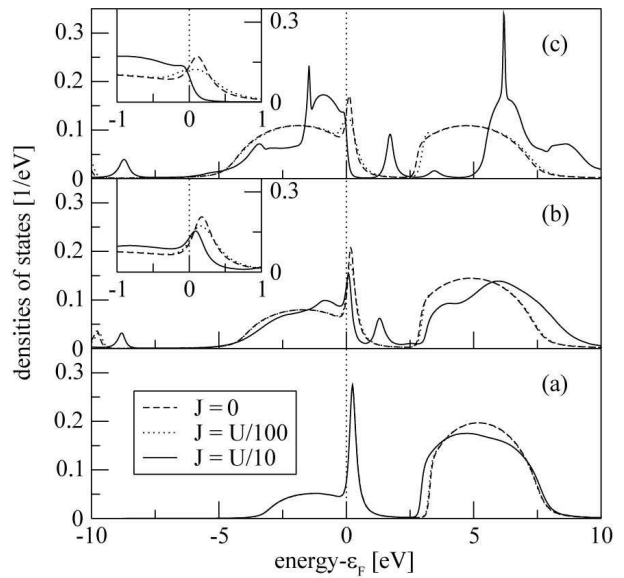}
\caption{Densities of states for various non-integer electronic
densities with Hund's exchange and for $U=6$~eV. (a) $n=0.9$,
(b) $n=1.9$ and (c) $n=2.9$. Insets show low energy excitations.}
\label{U6n_dop}
\end{center}
\end{figure}
%%%%%%%%%%%%%%%%%%%%%%%%%%%%%%%%%%%%%%%%%%%%%%%
In figure~\ref{U6n_dop}, the calculated densities of states  are displayed for various values of $n$~:
$n=0.9$, $n=1.9$ and $n=2.9$. They
correspond to  strongly
correlated metals close to the insulating state. The narrow
quasi-particle peak at the Fermi level is a signature of the large
electronic effective mass close to the MIT. For each $n$ and for
$U=6$~eV, the calculation has been done for three values of Hund's
parameter $J=0$, $U/100$ and $U/10$. The (a) part of the figure shows
that, for $n=0.9$, the effect of a finite $J$ value is
restricted to high energy structures of the spectrum.
This can be understood by considering the energies of the local states
listed in table~\ref{tab_energies}. For $n=0.9$, local states playing an
important role in the electronic dynamics are those with
occupations $0$, $1$ and $2$. Larger occupations of local states are
highly unlikely. The only $J$-dependent local energies are those for the 
two-fermion states $|3\rangle$, $|4\rangle$, $|5\rangle$ and
$|6\rangle$. Transitions from singly occupied state to these states
involve high energy excitations of the order of $U$. This $J$
influence restricted to high energy bands is specific to the singly
occupied multi-band Hubbard model.
Indeed, the physical situation is dramatically different for $n=1.9$ and
$n=2.9$ (part (b) and part (c) of figure~\ref{U6n_dop}). Here, a non
zero $J$ has a strong influence on high energy excitations as well
as close to the Fermi level. For low energy excitations an important
reduction of the  spectral weight is observed. This is
clearly visible in the figure insets and is in agreement with very
recent results at half filling for the doubly degenerate Hubbard
model with Hund's exchange.\cite{PB05} Here, we found that this
effect is also observed away from half filling. Besides, this
reduction is accompanied by the emergence of a multi-peak structure
around the Fermi level corresponding to the lifting of degeneracy of
multiparticle states in table~\ref{tab_energies}. This multi-peak
effect is stronger for $n=2.9$ which is consistent with the energy
interval $3J$ between the three-fermion local states of spin $3/2$ and
$1/2$. If we approach $n=3$, the peak at around $3J$
diminishes such that we are left with a gap of the order of $U$ at
$n=3$.

%%%%%%%%%%%%%%%%%%%%%%%%%%%%%%%%%%%%%%%%%%
\begin{figure}[hbtp]
%h=here, t=top, b=bottom, p=separate figure page
\begin{center}\leavevmode
\includegraphics[width=8cm]{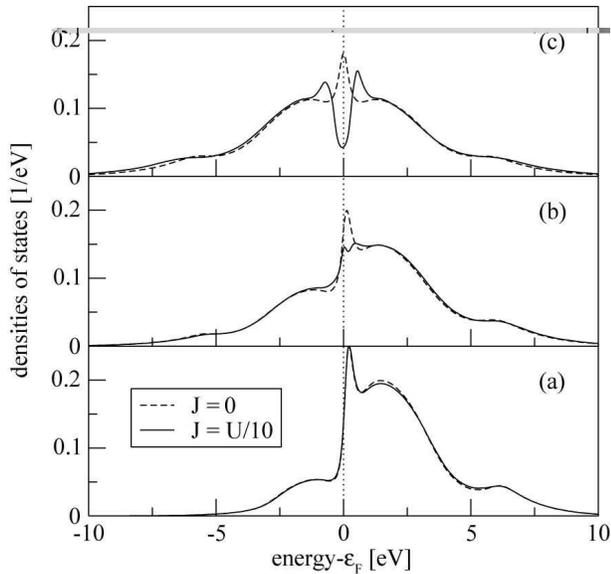}
\caption{Densities of states for various integer electronic
occupations with Hund's exchange and for $U=2$~eV. (a) $n=1$,
(b) $n=2$ and (c) $n=3$.} \label{U2n_int}
\end{center}
\end{figure}
%%%%%%%%%%%%%%%%%%%%%%%%%%%%%%%%%%%%%%%%%%%%%%%
The integer filling situation is examined in figure~\ref{U2n_int}.
The stability of the metallic state is ensured by the intermediate
value $U=2$~eV which is smaller than the critical $U_c$ of the MIT
for different integer fillings. $J$ takes values $0$ and $U/10$.
Here again, the effect of a non zero $J$ is very weak for $n=1$
(part (a)) and restricted to high energy excitations. A completely
different behavior occurs for $n=2$ and $n=3$. For $n=2$ (part (b)),
the introduction of a non vanishing $J=U/10$ is sufficient to
destroy the quasiparticle coherent peak, which is consistent with
the NRG results for the two-orbital Hubbard model.\cite{PB05} This
effect becomes stronger for the $n=3$ situation where a pseudo-gap
is observed. Such a structure with a pseudo-gap at the Fermi energy
is already present at half filling in the solution for the local
impurity model proposed in ref.~\onlinecite{PB05}. Here we show that this
critical role of $J$, leading to a pseudo-gap appearance, can be observed
away from half filling but the critical Coulomb and Hund parameters
are strongly dependent on $n$. This property has important
consequences on the nature of the different MIT in degenerate
systems. A multi-band Hubbard model displays a Mott transition at
all integer fillings. Nevertheless, by taking into account  the
Hund's exchange term, the different MIT present different low
energy dynamical behavior. It has been shown in previous studies
that the critical correlation strengths $U_c$ depend on $n$. For
instance, for a multi-orbital Hubbard model without Hund's coupling
and for a flat density of states, S. Florens and A.
Georges\cite{FG04} found that $U_c$ is proportional to $n(1-n)$. The
maximum critical coupling $U_c(n)$ for three orbitals is 
obtained for $n=3$, the half-filled situation, where orbital
fluctuations are largest. This result is consistent with previous
work, in particular M.J. Rozenberg\cite{R97} has found that
$U_c(2)>U_c(1)$ for the doubly degenerate orbital (with $J=0$)
Hubbard model by using Quantum Monte-Carlo calculations.

Concerning the influence of the Hund's coupling $J$ on the Mott
transition at $T=0$, an interesting generalization of the linearized
dynamical mean field theory has been proposed recently.\cite{OPB03}
In addition to a reduction of $U_c$ due to $J$, a qualitative change
in the nature of the $T=0$ Mott transition has been observed for any
finite $J$. The reduction of $U_c$ has been given in details only
for $n=M$ where $M$ is the degeneracy. Here we find important
differences between the half-filled case $n=M$ and the $n\neq M$
situation. Besides, our approach is able to explore dynamical properties
by calculating the full densities of states.

%%%%%%%%%%%%%%%%%%%%%%%%%%%%%%%%%%%%%%%%%%
\begin{figure}[hbtp]
%h=here, t=top, b=bottom, p=separate figure page
\begin{center}\leavevmode
\includegraphics[width=8cm]{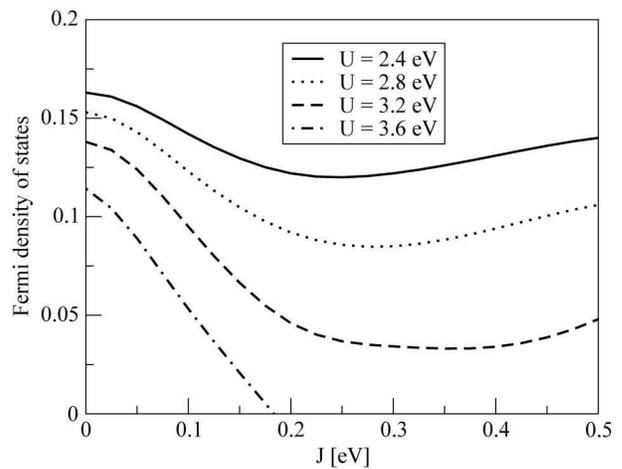}
\caption{Density of states at the Fermi level for
different  values of $U$ from $2.4$ to $3.6$, as a function of $J$
for $n=2$. Note that the system becomes
insulating for a critical $J$ if  $U$ is large enough.}
\label{n2_Jvar}
\end{center}
\end{figure}
\begin{figure}[hbtp]
%h=here, t=top, b=bottom, p=separate figure page
\begin{center}\leavevmode
\includegraphics[width=8cm]{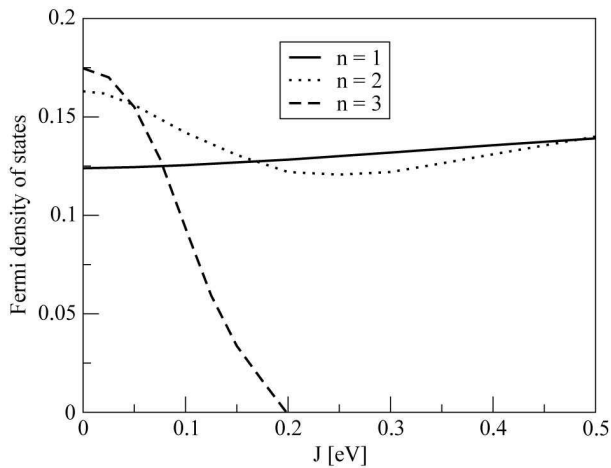}
\caption{Densities of states at $\varepsilon_F$ as a function of $J$ for
various integer occupations $n=1$, $2$ or $3$, and for $U=2.4$~eV.
Note that $J$ plays an important role at half filling $n=3$ but also
away from half filling for $n=2$.} \label{nvar_Jvar_U24}
\end{center}
\end{figure}

%%%%%%%%%%%%%%%%%%%%%%%%%%%%%%%%%%%%%%%%%%%%%%%
Figure~\ref{n2_Jvar} shows the $J$ influence on the
density of states at the Fermi level for a filling $n=2$ and for several $U$ values. 
A non-zero $J$ reduces low energy density of states.
Interestingly, for moderate $U$, there is a specific 
value of $J$, noted $J_m$ for
which the effect is the strongest. It is clear from
figure~\ref{n2_Jvar} that $J_m$ increases with $U$ and we have
approximately $J_m\approx U/10$. For large enough $U$, the increase of $J$ 
leads to a vanishing
spectral weight at the Fermi level. The resulting insulating phase is
driven by Hund's exchange. Note that for $n=2$, we found such an
insulating phase for $J_c\approx 0.2$~eV and only for $U>3.2$~eV. A
comparison with figure~\ref{nvar_Jvar_U24} shows the importance of
the band filling $n$.

%%%%%%%%%%%%%%%%%%%%%%%%%%%%%%%%%%%%%%%%%%
%%%%%%%%%%%%%%%%%%%%%%%%%%%%%%%%%%%%%%%%%%%%%%%
Indeed, in figure~\ref{nvar_Jvar_U24}, the dependence of the density
of states at $\varepsilon_F$ on $J$ is plotted for
$n=1$, $n=2$ and $n=3$, and for a constant $U=2.4$~eV. For $n=1$,
the influence of Hund's exchange is very weak. The only effect on
the low energy excitations is a light increase of spectral weight.
An appreciable decrease of the density of states at $\varepsilon_F$ is
found for $n=2$ and $U=2.4$~eV but the effect is much stronger for
the half-filled system $n=3$. Similarly to the $n=2$ situation, the
density of states at $\varepsilon_F$ goes to zero for a critical value
$J_c\approx 0.2$~eV but for $n=3$ this takes place at a lower 
repulsion value $U=2.4$~eV.

\section{Conclusion}
We have proposed an approach based on the dynamical mean field
theory which is able to describe important features of dynamical
electronic properties of the triply degenerate Hubbard model. The
self consistent local impurity problem is solved with a
generalization of the non-crossing approximation.
This impurity problem consists in a triply degenerate
local correlated site ($64$ local states) embedded
in an uncorrelated conduction band which is orbital dependent.
As expected, for $J=0$, we obtained a Mott metal insulator transition
for each integer filling namely $n=1,2,3,4,5$.

The influence of a Hund's exchange invariant under spin rotation 
has been studied in detail. In
agreement with previous works, we found that a non-zero $J$ implies
a reduction of the low energy spectral weight. This effect is
observed for integer as well as for non-integer fillings. In
addition, we established the occurrence of multi-peak effects in the
spectral function  at the metallic side of the MIT. These effects
give rise to structures with an energy scale $J$. An important
result of our approach is the very strong dependence of all these
effects on the filling. The destruction of low energy
states is much more pronounced at half filling than at other
fillings. Besides, for $n=1$, this effect is not visible.
Another interesting point is the occurrence of a particular value
$J_m$ for which the reduction is maximum, $J_m$ increasing with $U$.

It could be interesting to include in the present model different
bandwidths for different orbitals to investigate the role of Hund's
exchange in the so-called {\it orbital selective Mott 
transition}.\cite{KKR05}
Another possible perspective for this work is to understand how
disorder in real materials can affect the presented results
concerning Hund's exchange influence.
Finally, the pair hopping term between different orbitals (orbital-flip term) 
can be studied in the framework of our approach and will be the subject
of a forthcoming publication, as well as a more specific treatment
of realistic spectra.

We acknowledge valuable discussions with N. B. Perkins.

\end{document}